\documentclass[aip,jcp]{revtex4-1}

\usepackage[version=3]{mhchem} 
\usepackage{verbatim}
\usepackage{graphicx}
\usepackage{amsmath}
\usepackage{bm}

\begin{document}

\title{The proton momentum distribution in strongly H-bonded phases of water; a critical test of electrostatic models}
\author{C. J. \surname{Burnham}}
\email{christianjburnham@gmail.com}
\author{G. F. \surname{Reiter}}
\affiliation{Physics Department, University of Houston, Houston, Texas 77004}
\author{T. \surname{Hayashi}}
\author{S. \surname{Mukamel}}
\affiliation{Department of Chemistry, University of California, Irvine, California 92697-2025}
\author{R. L. \surname{Napoleon}}
\author{T. \surname{Keyes}}
\affiliation{Department of Chemistry, Boston University, Boston, Massachusetts 02215-2521}

\begin{abstract}
Water is often viewed as a collection of monomers interacting electrostatically with each other. We compare the water proton momentum distributions from recent neutron scattering data with those calculated from two electronic structure based models. We find that below 500 K 
the electrostatic models are not able to even qualitatively account for the 
sizable vibrational zero-point contribution to the enthalpy of vaporization.
This discrepancy is evidence that the change in the proton well upon solvation 
cannot be entirely explained by electrostatic effects alone.
\end{abstract}

\maketitle

\renewcommand \thesection{\Roman{section}}
\bibliographystyle{unsrt}
\pagestyle{myheadings}
\markright{The proton momentum distribution... \today}
\newcommand{\bra}[1]{\langle #1|}
\newcommand{\ket}[1]{|#1\rangle}
\newcommand{\braket}[2]{\langle #1|#2\rangle}

\section{INTRODUCTION}
A deep understanding of the physical and chemical properties of water in its liquid form 
is fundamental to the progress of 
atmospheric, biological and chemical sciences. Even though the properties of this 'elementary' molecule 
have been the subject of intense research for decades (and centuries), water has not yet yielded all its mysteries.

Fundamental to our understanding of the properties of water is a detailed description of the H-bond energetics and how these 
intermolecular bonds are modified according to differing environments. Considerable effort has gone into characterizing these bonds. \cite{Arunan1} In order to develop a microscopic picture, it is useful to regard the covalently bonded O-H oscillators as probes of the comparatively low frequency intermolecular (O-H--O) H-bonds in which they participate. 
Neutron Compton scattering is such a probe, providing a measurement of the momentum distribution of the protons in the water. \cite{Andreani1}

The momentum distribution, $n(p)$ is a statistical average over single particle distributions, with a shape determined by the shapes of the proton wells. The variations in $n(p)$ with environment are a direct result of changes 
in the proton potential energy surface(PES), which in bulk water will be due to the perturbative effect of the solvating water molecules on the intramolecular PES. 
Measurement of the momentum distribution thus provides direct access to the changes in the PES upon solvation. That there are significant changes is well known. 
The water OH stretch frequency in the bulk is redshifted by several hundred wavenumbers with respect to the gas-phase, indicating 
a significant softening in the proton PES by the intermolecular interactions in the bulk. Identifying the intermolecular forces responsible for this softening is still an open problem.

It seems reasonable to expect that intermolecular electrostatics accounts for much of the softening, but just how much is not known. 
Several studies ~\cite{corcelli1,schmidt1,ljungberg1} have convincingly demonstrated that there is a strong (though not necessarily linear) correlation between 
the intermolecular electric field and the OH stretch frequency in bulk water/ice. It remains an open question whether intermolecular 
electrostatics alone are sufficient to explain the softening in the OH stretch. This has proven difficult to answer, in part 
because of the lack of accurate (geometry dependent) electrostatic models for water, though a promising start has been recently made in that 
direction.~\cite{burnham9,mankoo3}

The first attempt to simulate the 
momentum distribution of protons in ice Ih was performed by Burnham \emph{et al}~\cite{burnham9} who 
developed a 
normal-mode path-integral molecular dynamics algorithm for evaluating the reduced one-body density 
matrix elements $\rho(\mathbf{r},\mathbf{r}')$. The Fourier transform of this quantity being the 
particle momentum distribution. 
Simulations were performed on an ice-like water cluster using empirical water models. All 
of the water models gave similar differences to experiment. In particular, the models were found to overestimate the kinetic energy contribution 
from the stretch modes, suggesting that the models overestimate the stiffness along the stretch direction

This study was followed by the study of Morrone, Car ~\emph{et al}~\cite{morrone1} who used a 'staging' based 
path-integral molecular dynamics technique with the flexible / non-polarizable SPCF-2 water model of 
Lobaugh and Voth~\cite{lobaugh1} 
to find the proton distribution in ice Ih, ambient liquid water and super-critical water at 673$^0$ K. 
Morrone ~\emph{et al} working with a simple modification of existing algorithms, used 
an approximation which appears to work very well for water 
in which $N_{mol}$ one body density matrices (on one proton per molecule) are calculated per simulation. Consequently, their algorithm is $N_{mol}$ times faster than 'serial molecule' calculations without this approximation.

The empirical water model used by Morrone ~\emph{et al} did not give any particular improvement in predictions for 
the momentum distributions over those of Burnham ~\emph{et al}. However, Morrone and Car have recently reported ~\cite{Morrone2}
a follow-up study, showing that momentum distributions calculated from a 
Car-Parrinello ~\cite{car1} molecular dynamics simulation using the BLYP functional gives much closer agreement with the 
experimental data for bulk ice and water, whereas none of the empirical models have come close 
to reproducing the accuracy of their Car-Parrinello based calculations for the proton momentum distributions in water/ice.

Though the Car-Parrinello results are impressive, it is important to note that almost all of the exchange correlation functionals (including BLYP) 
 being used in bulk-water simulations underestimate the frequency of the OH stretch band in both the bulk~\cite{silvestrelli2, tuckerman4} and gas phases
~\cite{Koch1, Campen1}. For instance, 
the bulk D$_2$O water/ice Car-Parrinello simulations of Chen \emph{et al} ~\cite{chen1} produce infrared spectra which underestimate the experimental OD 
stretch frequency by $\approx$ 300 cm$^{-1}$. The authors conclude that around 80 wavenumbers of this discrepancy is due to artefacts associated with 
the fictitious mass, leaving around 220 cm$^{-1}$ attributable to the approximate nature of the PBE functional used in the calculation. Unfortunately, 
quantization of the vibrational degrees of freedom is expected to increase the discrepancy, perhaps by another 50 cm$^{-1}$. 

The poor performance of commonly used functionals in reproducing the correct monomer distortion energies in bulk simulation of water, as compared to 
the much more accurate CCSD(T) predictions for the single molecule PES, was recently critically examined by Santra \emph{et al} ~\cite{santra1}. 
Consistent with previous studies, they found that both the BLYP~\cite{becke1, lee1} and PBE~\cite{perdew1} functionals significantly underestimate 
the stretch force constant. They also discovered that the PBE0 ~\cite{adamo1} hybrid functional does in fact seem capable of satisfactorily reproducing 
the variation in the PES along the OH stretch coordinate; at least for distorted monomers. It should be noted that the success of this functional 
was found to be at least partly due to fortuitous cancellation of errors in the exchange and correlation components (again compared 
to the CCSD(T) results). 

Most recently, Burnham \emph{et al}~\cite{burnham10} attempted to modify the electrostatics of the 
flexible/polarizable TTM2-F water model of Burnham \emph{et al} ~\cite{burnham5,fanourgakis1} to create a more realistic proton potential. 
They argued (based on the study by Hermansson ~\cite{hermansson1}) that a major factor in 
the softening of the OH stretch in the condensed phase is due to the geometry dependence of the 
OH bond polarizability. In nearly all polarizable water models the molecular polarizability is assumed independent 
of the molecular geometry. However electronic structure calculations demonstrate that the polarizability 
along the bond direction increases substantially as the OH stretch is elongated, which acts to 
soften the proton potential and also increases the bond dipole derivative in condensed phases. A 
parameterization of a water model that incorporated these effects was shown to produce a superior 
description of both the red-shift of the stretch band (reflecting a softening of the proton potential) 
and enhancement of the IR intensity (associated with the increased bond dipole-derivative) in 
agreement with experiment. However, this new model still overestimates the OH stretch, giving only a very 
minor improvement in the predicted momentum distribution with respect to experimental data.

The experimental momentum distribution has yet to be reproduced by any of the empirical models. In this 
work we calculate the momentum distribution from an electronic structure based-surface. 
This present study follows on from the work of Hayashi, Mukamel \emph{et al} ~\cite{hayashi1} who performed a computational study of the IR spectrum of HDO using a mixed QM/MM method. They first performed a simulation of the liquid using various empirical water models (polarizable and 
non-polarizable). The time-dependent electrostatic field about each molecule was then characterized 
by coefficients in a Taylor series expansion in displacements $\Delta \mathbf{r}$ about each molecule's center of 
charge. The electronic structure response to a given (non-uniform) field was then 
characterized by a second Taylor series, this time of the intramolecular potential energy surface 
(PES) and molecular dipole surface in vibrational normal-mode coordinates expanded in terms of field 
components. Combining the two series allowed for an electronic structure calculation of the 
(time-dependent) vibrational PES due to the intermolecular field from surrounding (empirically modeled) 
water molecules. The nuclear Schr\"odinger eqn was then solved on the resulting potential energy surface 
to find the vibrational eigenvalues and hence the $0\rightarrow 1$ OH stretch transitions, which can then be 
compared to the experimental IR spectra.  

It was shown by Hayashi, Mukamel ~\emph{et al} that their method gives an accurate representation of the position and 
line shape of the OH band in liquid HDO. However, in a subsequent study~\cite{Paarmann1} of H$_2$O, a scaling factor of 2.2 was 
required in the fields in order to bring the predicted and experimental OH frequency band into coincidence. Using 
the unscaled field distribution, the model will underestimate the size of the observed redshift of the bulk OH stretch frequency 
with respect to the gas phase. This is consistent with the hypothesis that whilst the redshift is strongly correlated 
with the intermolecular field, these fields are not strong enough by themselves to account for all or even nearly all of the complete 
redshift.

In this work, we use two different, but related models for calculating the momentum distribution in 
solvation. In both approaches electrostatic interactions are the only source of perturbation to the 
intramolecular PES of each water molecule. The results will be compared with experimental neutron scattering data for 
water at different state points from 300K to 673K. We find that the electrostatic interactions are insufficient to explain the softening of the 
PES in the denser phases of water. 

\section{MOMENTUM DISTRIBUTIONS}

In this secion we will summarize a few properties of nuclear momentum distribution functions.

\begin{figure}[h]
\begin{center}
\includegraphics[angle=90,scale=0.6]{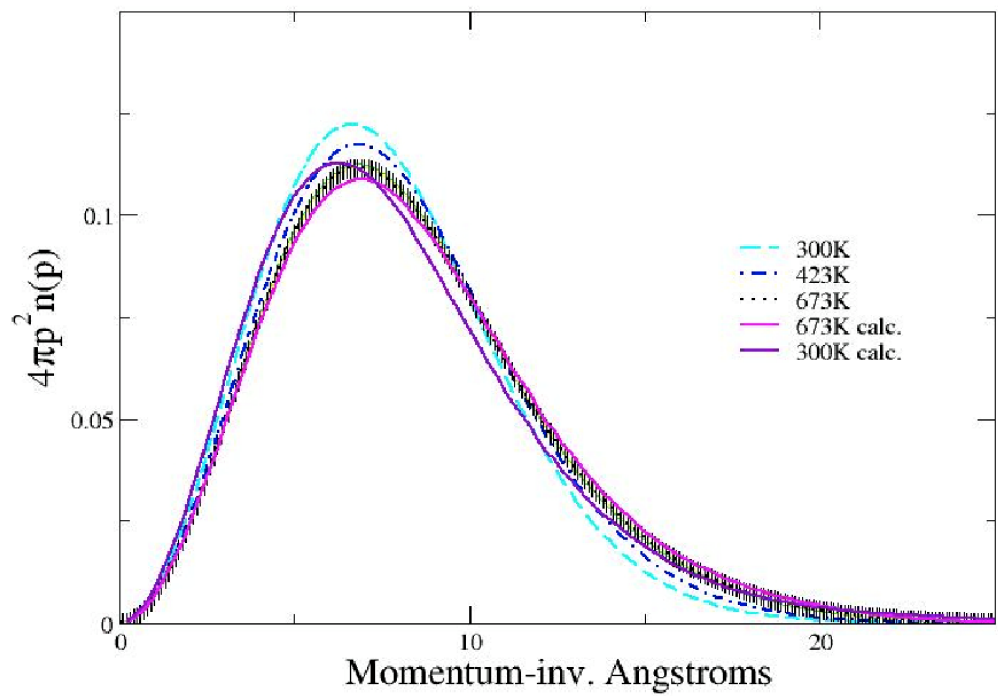}
\caption{Radial proton momentum distribution at several measured temperatures compared with path-integral 
molecular dynamics calculations using an extended TTM2-F model. Error bars on the lower temperature measurements have 
been omitted for clarity and are similar to those shown.  (Figure reproduced from ~\emph{Pantalei et al}\cite{pantalei1}).
}\label{fig3}
\end{center}
\end{figure}

The experimental momentum distributions at the various state points covered in this work are displayed in ~\ref{fig3}.

As motivation for the following, we have reproduced in ~\ref{fig3} the experimentally determined momentum distributions of \emph{Pantalei et al} 
for the various state points covered by this present study.  Note the figure also includes a comparison to our earlier path-integral based 
simulations which were performed using the TTM2-F water model of ~\emph{Burnham et al} ~\cite{burnham5,fanourgakis1}.  

Let the probability of a particle possesing a momentum in the range $\mathbf{p}..\mathbf{p}+d\mathbf{p}$ 
be $\rho(\mathbf{p})d^3\mathbf{p}$, where $\rho(\mathbf{p})$ is the momentum distribution function. Equivalently, 
in terms of the wave vector $\mathbf{k}=\mathbf{p}/\hbar$, we have 
$\rho'(\mathbf{k})=\hbar^3 \rho\left(\mathbf{p}=\hbar\mathbf{k}\right)$.

It is useful to consider the harmonic limit, in which the momentum distribution becomes a Gaussian ellipsoid 
$\rho^g(\mathbf{k})=\exp\left(-\mathbf{k}.\mathbf{T}.\mathbf{k}/2\right)/((2\pi)^{3/2}|\mathbf{T}|^{1/2})$, where 
the tensor $\mathbf{T}$ is defined by 
$\mathbf{T}=\sum_{\alpha=x,y,z} \hat{\bm{\alpha}}\hat{\bm{\alpha}}/\sigma^2_\alpha$, with 
$\hat{\bm{\alpha}}$ being the orthonormal ellipsoidal axes directions and 
$\sigma^2_\alpha=\left<k^2_\alpha\right>$ is the variance along the $\alpha^{th}$ axis of the ellipsoid. 

The observed $n(\mathbf{p})$ of protons in bulk water/ice is generally close to a (prolate) Gaussian and can be 
usefully characterized by the sigmas $\sigma_S > \sigma_{T1}\approx\sigma_{T2}$, where the \emph{stretch sigma} $\sigma_S$ 
is dominated by contributions from the OH stretch modes and the two transverse modes $\sigma_{T1},\sigma_{T2}$ are 
mostly due to motion in the bend and rotational modes of the waters. 

We will focus on the isotropically averaged distribution functions 
$n(k)=\left<\rho(\mathbf{k})\right>_{\phi,\theta}$. Note that the second moment of this function is related to the 
average kinetic energy per proton, via $\left<K.E.\right>=\hbar^2 \left<k^2\right>/2m_p=\left(\hbar^2/2m_p\right)\int{4\pi k^2n(k) dk}$, where $m_p$ is 
the proton mass.

For an arbitrary distribution (i.e. non Gaussian) we define the RMS sigma 
$\sigma_{RMS}=\sqrt{\left<\sigma^2\right>}$ from $\left<\sigma^2\right>=2m_p\left<K.E.\right>/3\hbar^2$, 
which is defined so that $\left<\sigma^2\right>=(1/3)\sum_{\alpha=x,y,z} \sigma_\alpha^2$ for Gaussian distributions.~\cite{conversion1}

The variance of the proton momentum distribution along the stretch direction is almost completely 
determined by the ground state vibrational wave-function and is of a fundamentally quantum nature. Thus 
quantum methods are required to obtain the distribution. 

In the ground state, the momentum distribution of the $n^{th}$ particle is given by

\begin{equation}\label{eqn1.5}
\rho_n(\mathbf{k}_n)=\frac{1}{\left(2\pi\right)^{3}}\int e^{-i\mathbf{S}_n.\mathbf{k}_n} \left<\hat{D}(\mathbf{S}_n)\right>d^3\mathbf{S}_n
\end{equation}

where $\hat{\mathbf{k}}_n=-i\partial/\partial \mathbf{r}_n$ is the momentum operator for the $n^{th}$ particle, i.e. 
and $\left<\hat{D}\left(\mathbf{S}_n\right)\right>$ is the thermal (or 
\emph{trace}) average of the displacement operator $\hat{D}(\mathbf{S}_n)=\exp\left(i\mathbf{S}_n.\hat{\mathbf{k}}_n\right)$.

The following strategy was used to obtain momentum distributions from the vibrational wavefunction. 
We solve the molecular wave-function in normal mode coordinates from which we evaluate the one-body density matrix elements $\rho_n\left(\mathbf{r}_n,\mathbf{r}_n'\right)=\bra{\mathbf{r}_n}
\left<\exp\left(-\beta \hat{H}\right)\right>_{m\ne n}\ket{\mathbf{r'}_n}$ where the inner triangular brackets 
denote a trace average of the full density matrix $\hat{\rho}(\beta)=\exp\left(-\beta \hat{H}\right)$ over 
all particles $m\ne n$. In our case, we are calculating properties for a single proton from a molecular 
wavefunction- thus the trace average is taken over the remaining H and O nuclei of the molecule. 

The trace average of the displacement operator is then given by the autocorrelation function

\begin{equation}\label{eqn1.6}
\begin{split}
\left<\hat{D}(\mathbf{S}_n)\right>=\int{\rho_n\left(\mathbf{r}_n-\frac{\mathbf{S}_n}{2},\mathbf{r}_n+\frac{\mathbf{S}_n}{2}\right)}d^3\mathbf{r}_n.
\end{split}
\end{equation}

Finally, from eqn.~\ref{eqn1.5}, the momentum distribution is obtained from Fourier transforming the 
above trace average back to $k$ space.

\section{EVALUATING THE VIBRATIONAL WAVE-FUNCTION}
We begin this section with a description of the two approaches we used for calculating the vibrational wavefunctions. Although the first 
of the two methods below does not (in this implementation) give excited state properties, there is practically no population in the vibrational mode 
excited states over the temperature range considered in this work ($<$1000 K) and calculation of ground state properties alone suffices.

The section continues with a discussion of harmonic corrections which are used to account for the fact that the nuclear wavefunction is solved 
in only a subspace of the complete set of molecular modes. Finally, we briefly desribe how the momentum distributions were obtained from the 
resulting harmonic-corrected wave-function eigenvectors. 

\subsection{Taylor Series approach}
In the first approach, hereafter referred to as the Taylor series approach, the ground state nuclear wave-function of a gas-phase molecule is calculated by
solving the vibrational Schr\"odinger eqn.in normal mode coordinates on the Born-Oppenheimer surface obtained from 
electronic structure calculations. The eigenvector describing the nuclear ground-state wavefunction is then Taylor 
series expanded as a function of external field and field derivatives in order to characterize the electrostatic response 
of the molecule. Field distributions in the condensed phases are taken from a molecular dynamics simulation in 
periodic boundary conditions using an empirical force field for water. Taking together the calculated field distributions and the field 
response as characterized by the Taylor series expansion, this method allows us to calculate the nuclear ground-state 
wavefunction of a single water molecule in the condensed phase.

The vibrational Schr\"odinger eqn. for a gas-phase molecule is solved in the space of its three vibrational modes (bend,
symmetric, antisymmetric stretch) using product basis states $\Pi_m\phi_{Am}(q_m)$, where $\phi_{Am}(q_m)$ is a Hermite basis 
function for the $m^{th}$ normal mode. The vibrational wavefunction for the $i^{th}$ energy level is then given by 

\begin{equation}\label{eqn1.1}
\psi^i(q_1,q_2,q_3)=
\sum_{A}C^i_A\phi_{A1}(q_1)\phi_{A2}(q_2)\phi_{A3}(q_3),
\end{equation}

where $C^i_A$ are the eigenvectors of the $i^{th}$ state, and the index $A$ labels the triplet 
$A=({A1,A2,A3})$. 

We next expand the electrostatic potential $\phi(\mathbf{r})$ in a Taylor series about the center of charge. Using the notation $r_\alpha=\hat{\bm{\alpha}}.\mathbf{r}$ and $E_{\alpha\beta}=-\partial^2 \phi/\partial r^X_{\alpha}\partial 
r^X_{\beta}$

\begin{equation}\label{eqn1.3}
\begin{split}
\phi(\mathbf{r}^X+\Delta\mathbf{r})-\phi(\mathbf{r}^X)=
-\sum_{\alpha}E_\alpha\Delta{r}_{\alpha}\\
-\frac{1}{2}\sum_{\alpha,\beta}E_{\alpha\beta}
\Delta{r}_{\alpha}\Delta{r}_{\beta}+...
\end{split}
\end{equation}

Finally, the eigenvector coefficients are also Taylor expanded, as a power series in the field terms

\begin{equation}\label{eqn1.4}
\begin{split}
C_A=C^0_A+\sum_j\frac{\partial C_A}{\partial E_j}\Delta{E_j}\\
+\frac{1}{2}\sum_{jk}\frac{\partial^2 C_A}{\partial E_j E_k}\Delta{E_j}\Delta{E_k}+...,
\end{split}
\end{equation}

where the $j$ index sums over all $\Delta E$ indices ($\Delta E_\alpha$,$\Delta E_{\alpha\beta}$ etc.). 
$\Delta E_j=E_j-E^0_j$ and $C^0_A$ indicates the value of $C_A$ at 
$\mathbf{E}=\left<\mathbf{E}\right>$ where the 
average is taken over all molecules. The electronic structure monomer field response is 
calculated with respect to a reference liquid geometry and $\left<\mathbf{E}\right>$ are the average field components 
in the liquid.

\subsection{Multipole-based approach}

In the second approach, the 
Born Oppenheimer surface of the gas-phase molecule is written as the sum of the intramolecular potential energy surface 
(as obtained in zero field) and the electrostatic energy of the charge distribution in an inhomogeneous external field. 
The electrostatic response of a molecule in a field is then characterized from the geometry dependent multipole and polarizability 
surfaces using the electronic-structure charge distribution. Next, a non-linear least-mean squares fit of a polarizable water-model 
is performed in order to best reproduce the geometry dependent electronic structure multipole and polarizability surfaces. 

In common with the Taylor-series method, the condensed phase molecular structures are taken from molecular dynamics calculations. 
Here the vibrational Schr\"odinger equation is solved on a potential energy surface using 
the above-mentioned least-mean-squares fit water-model to calculate both the response of the molecule to the fields and the fields themselves.

A six site water-model was used to fit the multipole surfaces. Three sites are located on the nuclear sites and
three are off-nuclear sites. Each nuclear site contains a permanent charge, a permanent dipole and a dipole-induction site. In 
addition two charge-sites are positioned symmetrically above and below the plane of the molecule along the molecular 
bisector. Finally, there is an off-nuclear induction site placed in the plane of the molecule and along the molecular bisector.

All electrostatic parameters in the water model (atomic charges and dipoles and location of off-nuclear sites) depend on the 
intramolecular geometry defined by $r_1,r_2,\theta$, where $r_1,r_2$ are the two OH 
distances and $\theta$ is half the HOH bond angle. Defining symmetry coordinates $S_1=(r_1-r_2)/\sqrt{2}$ and $S_2=(r_1+r_2)/\sqrt{2}$ each 
parameter is expanded in the power/Fourier series

\begin{equation}\label{eqn13.0}
\begin{split}
B_n(\Delta S_1,\Delta S_2,\theta)=\\
\sum_{i=0}^{i_{max}}\sum_{j=0}^i\sum_{k=0}^{k_{max}} b_{n,i,j,k} \left(\Delta S_1\right)^j \left(\Delta S_2\right)^{j-i} \cos\left(k \theta
+\alpha_n\right)
\end{split}
\end{equation}

where $B_n$ is the $n^{th}$ parameter and $\alpha_n$ is a constant. The expansion for each parameter is 
taken only over terms consistent with the molecular symmetry. 

The expansion terms are parameterized by minimizing the fitness function $A\left(\mathbf{b}\right)=\sum_m W_m \epsilon_m$ with respect to the 
expansion coefficients, where $\epsilon_m$ is the RMS error per tensor element over all elements of the rank $m$ (traceless) Cartesian multipole tensor 
and $W_m$ are weighting coefficients, with larger weights assigned to the lower rank tensor elements. 

The induction sites were fit to the gas-phase multipole polarizability surfaces, using a least mean squares fit to the gas-phase 
electronic-structure multipole moments in the presence of a small field in the $x,y$ and $z$ directions.

We found that the potential energy surface using the self-consistent multipoles becomes unstable, heading to -$\infty$ for extensions of 
the stretch coordinate by a few tenths of an Angstrom longer than the equilibrium value. This 
dipole-catastrophe occurs whenever the distance between interacting point-multipole induction sites gets too close. In order to remedy this unphysical behavior, the 
intermolecular dipole-dipole interactions were damped at short range based on the modified dipole tensor scheme of Thole~\cite{thole1}. 
Changes in the damping function parameters result in large effects on the calculated vibrational frequencies and in our implementation, these 
funcions were empirically fit to reproduce the frequency of the stretch band in the IR spectrum of ice. We find strong discrepancies in the calculated proton momentum distribution 
despite the fact that the vibrational spectrum is substantially correct. 

\subsection{Harmonic Corrections}

As noted above, in the Taylor series approach, eigenvectors are expanded as a function of 
$m=3$ vibrational modes (i.e. the two stretch modes and the bend), whereas those in the multipole based approach 
are expanded in three vibrational and three rotational modes, i.e. $m=6$ modes in full. In order to 
account for the fact that both approaches use only a subset of the total complement of nine degrees of freedom 
per molecule, additional harmonic corrections are made to account for the ($9-m$) modes missing from each 
calculation. The momentum distribution is calculated from the Fourier transform of the trace average 
of the density operator over the included anharmonic degrees of freedom multiplied by a Gaussian correction accounting for displacement along the 
missing molecular modes:

\begin{equation}
\left<\hat{D}(\mathbf{S}_n)\right>=\left<\hat{D}(\mathbf{S}_n)\right>_{A}\exp\left(-\frac{\mathbf{S}_n.\mathbf{T}_h^{-1}.\mathbf{S}_n}{2}\right)
\end{equation}

where $<...>_A$ indicates an average over the included anharmonic degrees of freedom and $\mathbf{T}_h^{-1}$ is a temperature dependent tensor
obtained from a harmonic analysis in which estimates were used for the harmonic frequencies of the missing modes. For the multipole model the 
tensor includes contributions from the three translational modes only, with the harmonic frequencies assigned to be $\omega_{trans}=$250cm$^{-1}$. 
In the Taylor-series based approach, only the vibrational degrees of freedom are accounted for and so harmonic corrections were required for 
both the three librational and three translational modes. For this model, estimates for the (temperature and density dependent) 
harmonic rotational frequencies were taken from calculated IR spectra using the multipole-based model.

\subsection{Calculation of the momentum distribution}

Finally, the momentum distribution was calculated from the eigenvectors of the vibrational wavefunction as follows.

The autocorrelation function of the single particle density matrix (see eqn.~\ref{eqn1.6}) was calculated in the harmonic oscillator basis using 
the known relation for the autocorrelation of a Hermite function in terms of associated 
Laguerre polynomials, e.g. see ref. ~\cite[Appendix B]{cahill1}. The resulting autocorrelation was then numerically 
Fourier transformed to give the single particle momentum distribution (see eqn.~\ref{eqn1.5}) and the final momentum distribution is taken to be 
the average over single particle momentum distributions.

\section{MODEL PARAMATERIZATION AND SIMULATION DETAILS}

All simulations used 128 molecules in periodic boundary conditions at experimental densities. Momentum distributions in the bulk were calculated 
using intermolecular geometries obtained from 'snapshots' of molecular dynamics path integral simulations in periodic boundary conditions using the TTMF-4 water model. As has been previously shown,~\cite{pantalei1} this model gives reasonable agreement with 
the experimentally determined structures over the temperature/density range to be covered in this present work. 

The path-integral simulations were used only to generate reasonable configurations representative of equilibrium 
structures and were not particularly computationally intensive, with the equilibration 
requiring only a few hours of CPU time per state point. 

\subsection{The Taylor-series approach}
This model was parameterized using MP2/6-31+G(d,p) electronic structure data.

The Taylor series approach requires intermolecular field distributions provided from an electrostatic model of the solvent molecules. We chose the 
TIP3P-F water model for this purpose because it was found to give similar field distributions to those used in the original study of Hayashi, Mukamel ~\emph{et al}. The intermolecular electrostatics of this model are obtained using static charges placed on the nuclear sites, 
with $q_O=-2q_H=-0.834 |e|$. 

In determining the electronic-structure electrostatic response the local field at each expansion center
was expanded up to fourth order (i.e. terms including $E_{\alpha\beta\gamma\delta}$). 
The wave-function expansion coefficients $C_A$ were expanded up to second order in 
$\Delta E_j$ (see eqn.~\ref{eqn1.4}). A total of 35 $C_A$ coefficients ($A1+A2+A3 \le 4$) were 
used in the expansion of the ground state wavefunction of eqn.~\ref{eqn1.1}. 

The electric fields were calculated via standard Ewald sum expressions. Gradients 
and higher order derivatives of the field were calculated using a finite difference method using as 
input the field evaluated at grid points around the field expansion center.

\subsection{The multipole based approach}
The multipole-based model was parameterized using RMP4/aug-cc-pvdz electronic structure data obtained using the 
Gaussian 03 code.~\cite{gaussian1}

The parameters were expanded up to a $6^{th}$ order polynomial ($i_{max}=6$) in $\Delta S_1,\Delta S_2$, with six Fourier 
coefficients over $\theta$, fit to reproduce ~1500 geometries covering the range $90^0\leq\theta_{HOH}\leq 130^0$ and 
$90^0\leq r_{H1},r_{H2}\leq 130^0$.

The fit resulted in accurate reproduction of the geometry-dependent electronic structure multipole surfaces up to rank-3 (octapole). The 
parameter set and FORTRAN code to run the model is available by request from the corresponding author.

The polarizabile multipoles on each molecule are solved iteratively for each different 
arrangement of nuclear coordinates, ensuring that the system remains at all times on the electrostatic self-consistent surface. 

Obtaining self-consistent multipoles across all molecules makes this approach far more expensive than the Taylor-series approach. 
In order to reduce costs, a simple real-space spherical cut-off was used for calculating electrostatic interactions, with a radius equal 
to half the shortest axis of the periodic simulation cell. 

The multipole model uses Partridge and Schwenke's accurate intramolecular surface~\cite{partridge1} for each monomer. Thus, 
the multipole model becomes identical to the Partridge Schwenke model in the gas-phase.

The wavefunction was calculated from scanning over three vibrational and three rotational modes per molecule. Performing a 
full 6-dimensional scan would be quite slow and so following Jung and Gerber~\cite{jung1}, 
a many-body approximation was employed in which only two body terms are retained, 
reducing the calculation to a more managable set of $6\times5/2=15$ 
scans per molecule. The wavefunction is then solved for by first calculating the vibrational self-consistent 
field (VSCF) solution~\cite{bowman1} in the 6D space. This results in a product wavefunction 
$\psi^{VSCF}_\mathbf{m}(q_1,...,q_6)=\phi_{m_1}(q_1)\phi_{m_2}(q_2)...\phi_{m_6}(q_6)$ with $\psi^{VSCF}_\mathbf{0}$ giving 
the optimal single product ground state. The final wavefunction is then solved using a configuration interaction 
in which the reduced Hamiltonian $H_\mathbf{m,m'}=\left<\mathbf{m}|H|\mathbf{m'}\right>$ is diagonalized,
where $H_\mathbf{m,m'}$ is formed from states $\mathbf{m}$ with $\sum_i m_i\leq 4$ leading to matrix sizes of $\approx 6500\times 6500$. Altogether it 
takes $\approx 40$ minutes to calculate the momentum distribution for 
one particle, with the diagonalization stage being by far the most 
time-consuming part of the calculation.

\section{RESULTS}

\begin{figure}[h]
\begin{center}
\includegraphics[scale=0.8]{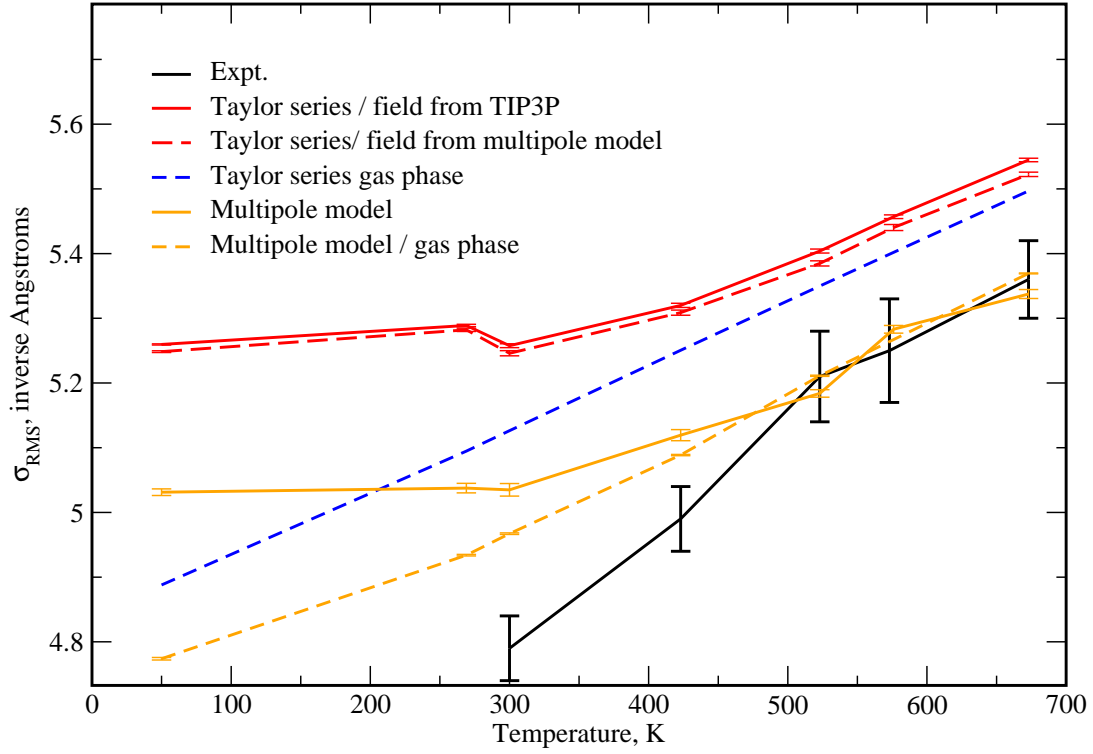}
\caption{RMS values of the observed and calculated proton momentum distribution width. These widths give directly the kinetic energy of the protons. The results are shown 
as a function of temperature, with densities 0.9 gcm$^{-3}$ at 50 and 269 K 1 gcm$^{-3}$ at 300 K, 0.9 gcm$^{-3}$ at 423 K, 0.8 gcm$^{-3}$ at 523 K, 0.7 gcm$^{-3}$ at 573 K and 673 K. (Data reproduced from \emph{Pantalei et al.}\cite{pantalei1}) The curve labeled gas phase is calculated by including the temperature dependence of classical rotation and translation modes in the calculation of the total energy in addition to the kinetic energy calculated for the vibrational modes of the isolated molecule. 
}\label{fig1}
\end{center}
\end{figure}

The calculated and experimental~\cite{pantalei1} RMS sigmas $\sigma_{RMS}$ for the set of temperature/density state points are displayed in ~\ref{fig1} as a function of temperature. At each temperature value we have also plotted calculated values for the gas-phase momentum distributions.

The bulk values for $\sigma_{RMS}$ are seen to approach the corresponding gas-phase values for 
$T\ge 500$ K. In 
this high temperature region there appears to be practically no difference in the kinetic energy between the bulk and gas-phases. 

The observed approximate linear dependence with temperature of the predicted gas-phase sigmas is due to the $k_BT/2$ per degree of freedom 
kinetic energy in the near-classical translational and rotational modes. In the temperature range under study the gas-phase 
vibrational modes are nearly completely frozen out, having virtually no contribution to the temperature dependence. 

Though the Taylor series and multipole-based approaches predict quite different absolute values 
for $\sigma_{RMS}$, the two show very similar trends 
as a function of temperature. Most of the difference between models appears to be due to their different gas-phase asymptotes, 
with the Taylor-series approach predicting a slightly larger value for the gas phase $\sigma_{RMS}$ at each temperature. Presumably 
this is due to the different intramolecular surfaces used in the two models, with the gas-phase asymptote of the Taylor series model 
being given by MP2/6-31+G(d,p) electronic structure vs. the somewhat more accurate Partridge-Schwenke gas-phase surface incorporated into 
the multipole based model.

The multipole-based approach is in excellent agreement with the experimentally determined sigmas at high temperatures, T$\geq 500$ K. 
As the temperature is lowered, the curves diverge, with the experimental $\sigma_{RMS}$ showing a steeper drop  
than predicted by either of the simulation approaches. At 300 K the discrepancy between the 
predicted and observed kinetic energies is 0.35 kcal/mol per proton, or 0.7 kcal/mol per molecule. This discrepancy 
between simulation and experiment accounts for $\approx7\%$ of the 9.7 kcal/mol per molecule enthalpy of vaporization, 
$\Delta H=H^{(g)}-H^{(s)}$ at 300K. 

Furthermore, the simulations predict the wrong sign for the kinetic energy contribution to $\Delta H$. Assuming that the 
gas-phase kinetic energies calculated using the high-quality Partridge-Shwenke intramolecular surface are correct, the 
experimental data shows that below 500 K, the kinetic energy per proton in the bulk lies below the gas-phase values (at the same 
temperature), whereas simulation predicts the opposite behavior, with the average kinetic energy per proton in the bulk 
becoming \emph{larger} than the corresponding gas-phase values at lower temperatures. 

The increase in the simulated kinetic energy is due to hindered rotations in the bulk having 
a larger kinetic energy than in the gas-phase, where the molecules act as free rotors. A similar contribution is expected to 
be present in the experimental data, but this is evidently more than cancelled out by the lowering in kinetic energy due to the softening in the 
stretch modes.

The experimentally observed reduction in $\sigma_{RMS}$ over the correspoinding values in the gas-phase indicates that the effective proton potential is substantially softened at low temperatures; a softening manifestly not reproduced 
by the electrostatic models. 

\begin{figure}[h]
\begin{center}
\includegraphics[scale=0.8]{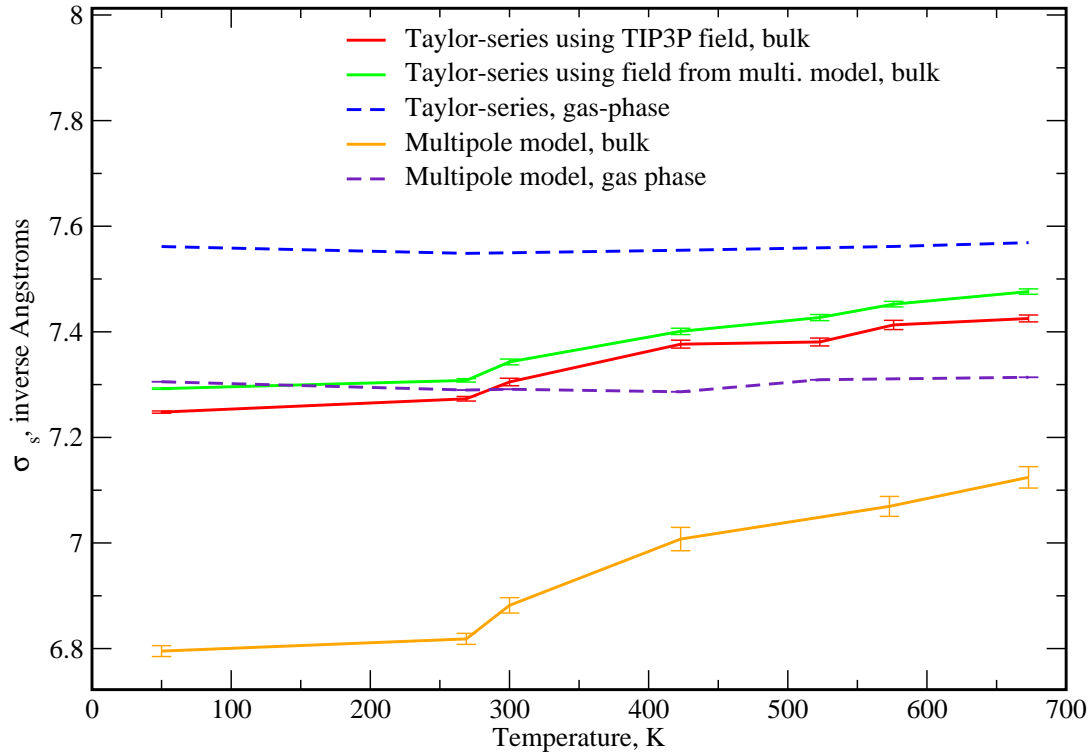}
\caption{The stretch sigma values(momentum witdths) of the calculated proton momentum distributions at the same temperature/density 
points as in Fig 1.}\label{fig2} 
\end{center}
\end{figure}

Experiment has been previously shown~\cite{pantalei1} that most of the reduction in the proton mean kinetic energy is 
due to a lowering of the stretch sigma ($\sigma_s$) values. ~\ref{fig2} shows simulation results for this sigma as a function 
of temperature, extracted from the calculated momentum distribution using the multipole and Taylor-series methods. The individual sigmas were then 
obtained from the tensor of second moments $M_{\alpha\beta}=\left<k_\alpha k_\beta\right>$, which when diagonalized gives eigenvectors $\hat{\alpha}$ 
with corresponding eigenvalues $\sigma^2_\alpha$.

Again, though there is a sizable offset between values calculated from the multipole and Taylor-series approaches, both approaches 
result in near identical trends as a function of temperature. The simulation shows a modest, but definite decrease in 
the stretch sigmas as the temperature is lowered. This reduction is due to a softening of the proton potential, in turn caused by 
the increasing strength of the intermolecular fields at lower temperatures and higher densities.

\section{SUMMARY AND CONCLUSIONS}
In this work we tested how accurately state of the art electrostatic models for water, which perform well 
for infrared spectra can reproduce the observed neutron scattering 
data for the proton kinetic energy and momentum distribution in bulk water. Both models were parameterized to reproduce high 
quality electronic structure data for the geometry dependent monomer electrostatics. However, two quite different modeling 
approaches were used in order to generate the nuclear wave-function from which the momentum distributions were calculated. 

Although the two electrostatic models give qualitively similar results for trends in the kinetic-energy and momentum widths, neither was able to fully reproduce the observed experimental trends. 
In contrast with experiment, both models predict the kinetic energy 
per proton increases in the strong H-bonding region. Even sophisticated electrostatic models for water, 
such as the ones used in this work are unable to reproduce the correct sign in the proton kinetic energy contribution to the 
enthalpy change between the gas and condensed phases at the same temperature.

These results have important implications for all simulations requiring accurate bulk properties of water. 
For a simulation to reproduce the experimental enthalpy difference (for the right reasons), it must take into account the zero-point 
changes associated with the vibrational degrees of freedom. Several recent studies have examined the thermodynamic properties of water 
using rigid-body path-integral calculations.~\cite{mcbride1,hernandez1,hernandez2,hernandez3,hernandez4} Though cheaper than a fully flexible path-integral simulation, such calculations obviously 
cannot account for the relatively sizeable changes discussed in this work associated in the proton kinetic energy contribution to 
the enthalpies.  In this regard it is also worth mentioning the recent simulations of Hyeon-Deuk and Ando,~\cite{hyeon1} who 
performed a wavepacket dynamics simulation of bulk liquid water using a simple empirical flexible water model.  They observed significant 
correlations between the wave-packet width of the proton with its H-bond coordination number, further confirming that quantization 
vibrational modes will affect the properties of the H-bond network.

Even larger discrepancies than those described above between the electrostatically interacting monomer model and experimental momentum distributions are found in water in confined spaces. Neutron scattering studies from ~\emph{Kolesnikov, Reiter et al} ~\cite{kolesnikov1,reiter3} 
has shown that water in carbon nanotubes displays a very sizeable(~30$\%$) reduction in the zero-point kinetic energy of the protons relative to bulk water, consistent with the anomalously large Debye Waller factors also observed. Measurements of the momentum distribution in water confined in xerogel by ~\emph{Garbuio et al}\cite{garbuio1} show, on the other hand, large increases in the kinetic energy of the protons in the water near the surface, also unexplainable with the electrostatically interacting monomer model. 

Finally, consider the possible reasons why the electrostatic models fail to predict the proton kinetic energy changes in 
the bulk. Given that the proton momentum distribution is a function of the vibrational potential energy surface, we are led to conclude 
that the intermolecular perturbation on the proton potential in bulk phases cannot be satisfactorily modeled using classical 
electrostatics alone. As evidenced by comparison with experiment, there appears to be an additional softening of the proton potential upon H-bond 
formation, beyond that accounted for by intermolecular electrostatics.

Why do the electrostatic models underestimate the required softening in the OH stretch? The most probable explanation is that these 
models either completely neglect or at best provide only a crude estimate of intermolecular overlap effects between H-bonded molecules, i.e. in a molecular orbital 
description of the wavefunction, the orbitals of the donor and acceptor molecules have a non-negligable overlap, resulting in a total wavefunction that cannot be 
accurately approximated as the product of single-molecule contributions. 

Overlap effects can be usefully subdivided into various contributions. First, there is charge-overlap/penetration term. The intermolecular 
energy of overlapping charge distributions differs from that predicted using traceless multipole-multipole interactions. Charge overlap always makes the 
interaction energy more positive with respect to the traceless multipole approximation (the difference in both the e-e and e-Z terms are also separately 
positive).

Second, there is the exchange contribution arising from the Pauli exclusion principal. The intermolecular contribution to the exchange energy is 
generally positive for an interaction between closed shell molecules. 

Lastly, there is electron correlation, which can be defined as the difference between the exact solution to the electronic Hamiltonian with 
respect to the self-consistent field solution obtained using the Hartree Fock method. The correlation contribution causes modifications 
to both the electronic kinetic and potential ($V_{e-e}$ and $V_{e-Z}$) expectation values, producing an overall reduction in the total energy. The 
intermolecular contribution is also negative and therefore correlation is the only one of the overlap effects expected to reduce the energy. 
Also, given that all of the overlap effects tend to increase with increasing overlap, the correlation term should result in an overall softening in the 
stretch PES of protons participating in H-bonds.

The Taylor-series approach as implemented in this work neglects all overlap effects. The multipole-based approach however incorporates 
the aforementioned short-range damping term in the fields in order to prevent over-polarization of the induction sites. Given that the damping functions 
attenuate the short-range (electrostatic) interactions, it seems plausible that a suitable parameterization of these terms can help to at least 
partly mimic the effects of overlap on the PES. We should not expect that such a simple modification to the PES can provide anything better 
than a crude approximation to the actual overlap effects present in the intermolecular interaction. Indeed, we have seen that even 
when damping functions are included, the multipole model is not accurate enough to give qualitative agreement with the observed kinetic energies.

It may be very difficult to modify existing electrostatic based models to account for the intermolecular overlap effects on the OH stretch PES 
in order that, upon quantization of the nuclear coordinates, they be capable of reproducing the observed kinetic energy changes. As things stand, 
it is doubtful whether any of the existing models are accurate enough to even qualitatively account for the environmentally dependent kinetic 
energy changes in the vibrational modes. Nevertheless, comparison of results against the experimentally determined proton momentum 
distributions provides an invaluable benchmark of both empirical models and exchange-correlation functionals, which should help in the effort to create more accurate simulations where water is involved and further reveal the nature of the H-bond.

\section{Acknowledgements}
C. J. Burnham and G. F. Reiter acknowledge support by the DOE, Office of Basic Energy Sciences under Contract No. DE-FG02-03ER46078.

S. Mukamel gratefully acknowledges the support of the National Institutes of 
Health (Grant GM59230) and the National Science Foundation (Grant CHE-
0745892)

Acknowledgment is made to the donors of the American Chemical Society Petroleum Research Fund for partial support of this research by 
T. Keyes and R. Napoleon. T. Keyes acknowledges support of the National Science Foundation (Grant CHE 0848427).

\clearpage

\clearpage

\clearpage

\bibliography{newmodel}

\end{document}